\documentclass[11pt,preprint]{aastex}
\newcommand{\etal}{{\it et~al.}}

\begin{document}

\title{The Euphrosyne family's contribution to the low albedo near-Earth asteroids}

\author{Joseph R. Masiero\altaffilmark{1}, V. Carruba\altaffilmark{2}, A. Mainzer\altaffilmark{1}, J. M. Bauer\altaffilmark{1}, C. Nugent\altaffilmark{3}}

\altaffiltext{1}{Jet Propulsion Laboratory/Caltech, 4800 Oak Grove Dr., MS 183-601, Pasadena, CA 91109, {\it Joseph.Masiero@jpl.nasa.gov, amainzer@jpl.nasa.gov, bauer@scn.jpl.nasa.gov}}
\altaffiltext{2}{Departamento de Matem\'{a}tica, Universidade Estadual Paulista, Av. Dr. Ariberto Pereira da Cunha 333., CEP 12516-410, Guaratinguet\'{a}, SP, Brazil {\it vcarruba@feg.unesp.br}}
\altaffiltext{3}{Infrared Processing and Analysis Center, Caltech, Mail Code 100-22, 770 South Wilson Avenue, Pasadena, CA 91125 {\it cnugent@ipac.caltech.edu}}

\begin{abstract}

The Euphrosyne asteroid family is uniquely situated at high
inclination in the outer Main Belt, bisected by the $\nu_6$ secular
resonance.  This large, low albedo family may thus be an important
contributor to specific subpopulations of the near-Earth objects.  We
present simulations of the orbital evolution of Euphrosyne family
members from the time of breakup to the present day, focusing on those
members that move into near-Earth orbits.  We find that family members
typically evolve into a specific region of orbital element-space, with
semimajor axes near $\sim3~$AU, high inclinations, very large
eccentricities, and Tisserand parameters similar to Jupiter family
comets.  Filtering all known NEOs with our derived orbital element
limits, we find that the population of candidate objects is
significantly lower in albedo than the overall NEO population, although
many of our candidates are also darker than the Euphrosyne family, and
may have properties more similar to comet nuclei.  Followup
characterization of these candidates will enable us to compare them to
known family properties, and confirm which ones originated with the
breakup of (31) Euphrosyne.

\end{abstract}

\section{Introduction}

\begin{figure}[ht]
\begin{center}
\includegraphics[scale=0.55]{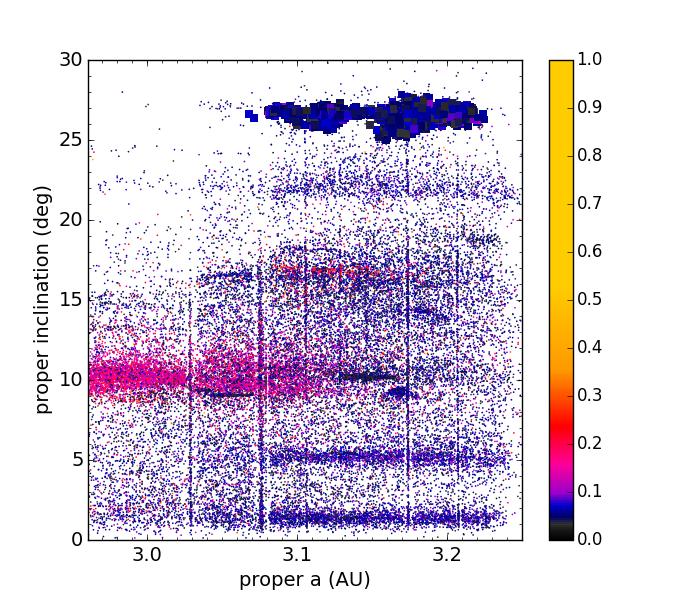}
\caption{Proper orbital inclination vs proper semimajor axis for all
  asteroids in the outer Main Belt (dots) and Euphrosyne family
  members (large squares).  Colors indicate visible geometric albedo
  following the colorbar. Vertical features are a result of weak
  resonances and a feature common to synthetic proper orbital element
  catalogs.  Horizontal features are asteroid families.  Orbital
  elements, albedos, and family identifications taken from
  \citet{masiero13}.}
\label{fig.orbs}
\end{center}
\end{figure}

The Euphrosyne asteroid family occupies a unique place in orbital
element space among families, located in the outer Main Belt at very
high inclination, as shown in Figure~\ref{fig.orbs}.  It is also the
only asteroid family that is bisected by the $\nu_6$ secular orbital
resonance.  The inner-most portion of this resonance (semimajor
axis$<2.5$ AU) has been shown to be a primary mechanism for moving
asteroids onto near-Earth orbits with relatively long lifetimes
\citep{bottke02}.  Conversely, \citet{bottke02} found that the Jupiter
mean motion resonances tended to quickly move objects onto
Jupiter-crossing orbits that lead to an impact or ejection from the
Solar system.

The largest remnant of this family, (31) Euphrosyne, is one of the
largest asteroids in the Main Belt, with a diameter of
$D=260\pm12~$km, an optical albedo of $p_V=0.05\pm0.01$
\citep{masiero14}, a bulk density of $\rho=1.2\pm0.6~$g cm$^{-3}$
\citep{carry12}, and a taxonomic classification of Cb
\citep{bus02tax}.  The Euphrosyne family is one of the most numerous
families, with approximately 1000 associated members
\citep{nesvorny12,milani14}.  The family has an average albedo of
$p_V=0.056\pm0.016$, very close to that of the largest remnant, and an
anomalously steep present day cumulative size frequency distribution
of $N(>D) \propto D^{-4.4}$ \citep{masiero13}.  As one of the few
large, low-albedo families near the $\nu_6$ resonance, Euphrosyne may
be one of the important contributors to the low albedo component of
the Near-Earth object (NEO) population \citep[cf.][]{mainzer11neo}.

In this work, we investigate the potential contribution of the
Euphrosyne family to the NEO population over the predicted $\sim700$
Myr age of the family \citep{carruba14}.  We quantify the most
likely region of orbital-element phase space that these family
escapees would be found in, and search the list of known NEOs for
candidate objects related to the family.  We also produce a shortlist
of followup candidates that have a high likelihood of being
compositionally related to the Euphrosyne family.

\section{Simulations}
\label{sec.sims}

To test the contribution of the Euphrosyne family to the near-Earth
population, we performed simulations of the orbital evolution of the
small asteroids created by the catastrophic impact event that formed
the observed family.  We use the SWIFT-RMVSY symplectic numerical
integrator \citep{holman93,levison94,broz06} to follow the evolution of these
test particles under both gravity and the Yarkovsky effect
\citep{bottke06}.  We performed seven independent simulations each
consisting of 4000 massless test particles, for a total of 28000
simulated family members.  For massive bodies in each simulation we
used the Sun, Mercury, Venus, the Earth-Moon barycenter, Mars,
Jupiter, and Saturn.  Test simulations carried out using all planets
as massive bodies show identical results to the restricted case; thus,
Uranus and Neptune were excluded to reduce run time of the
simulations.  Positions for the massive bodies were initialized to
their computed location on MJD$=57000$ from the JPL DE405 planetary
ephemerides file\footnote{\it http://ssd.jpl.nasa.gov/?planet\_eph\_export}.

Family members were initialized at the position of (31) Euphrosyne on
MJD$=57000$ and given a randomized velocity relative to the parent
body.  The magnitude of the velocity was scaled inversely
proportionally to the diameter of the test particle \citep{vok06},
with a characteristic velocity of $130~$m/s for an object with a
diameter $D=5~$km, following the best fit breakup velocity found by
\citet{carruba14}.  A secondary phase space minimum is present for a
significantly lower characteristic velocity of $\sim20~$m/s
\citep{carruba15}, so we also simulated a set of test cases with a
characteristic velocity of $20~$m/s to look for potential differences
in the final result that could result from a variation in this initial
condition.

The physical properties assumed for the test particles, which play an
important role in the strength of the Yarkovsky effect, were drawn
from the overall characteristics of the family.  The geometric albedo
of each object was assumed to be $p_V=0.056$, matching the mean for
the family given by \citet{masiero13}.  Bulk and surface densities
were assumed to be $1200~$kg/m$^{3}$, based on the measured values for
the largest remnant as well as for asteroids of similar spectral
taxonomic type \citep{carry12}.  As full thermophysical models of (31)
Euphrosyne or Euphrosyne family members are not available, we assume
nominal thermophysical parameter values for the simulated particles
based on previous research \citep{vok06}: thermal capacity
$C_p=680~$J/kg/K, emissivity $\epsilon=0.9$, and thermal conductivity
$K=0.01~$W/m/K.  The evolution of our test particles is dominated by
secular resonances and other gravitational effects in this case, and
Yarkovsky orbital evolution is not required to inject particles into
these resonances, thus the uncertainty in these thermal parameter
values does not dominate the simulation results.

As discussed by \citet{carruba14}, the slope of the family size
frequency distribution changes as the family evolves.  Thus, sizes for
simulated particles were generated to fulfill a cumulative size
frequency distribution slope of $\alpha=-3.6$, based on the
best-fitting initial value from \citet{carruba14} over the range from
$1~$km to $20~$km.  Test particles were given random initial spin
states, which evolve under YORP and collisional re-randomization as
described in \citet{broz06}.  \citet{bottke15} introduce a new method
for simulating YORP evolution they refer to as ``stochastic'' YORP,
however given the proximity of the family to the $\nu_6$ resonance,
this new effect is not expected to significantly alter the results as
compared to the older YORP simulation technique employed here.

Simulations were run forward for $700~$Myr, using $3~$day time steps
to ensure that minimal errors would be induced in the orbit of Mercury
or any test particle with a small perihelion distance.  We recorded
the orbit of each particle every 10 kyr over the entire simulation,
and selected all objects with perihelia $q<1.3~$AU, which would be
classified as NEOs.  We then tracked the orbital elements of each
simulated NEO to look for characteristic values that would allow us to
search the real NEO population for potential Euphrosyne family
members.

\section{Results}

\begin{figure}[ht]
\begin{center}
\includegraphics[scale=0.55]{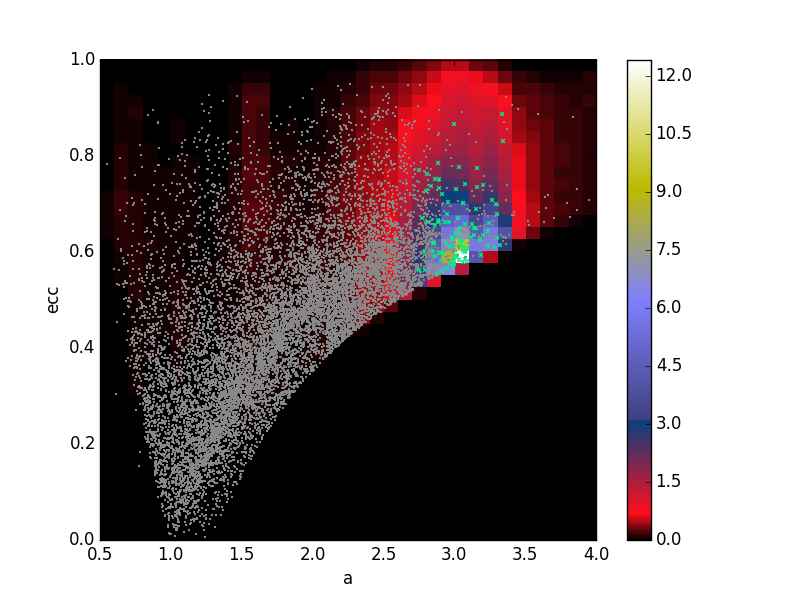}
\caption{Density map of orbital elements of all simulated Euphrosyne
  family members entering the near-Earth population. Bins of semimajor
  axis (a) and eccentricity (ecc) are colored according to the
  probability density shown in the colorbar.  Overplotted as points
  are all known NEOs in grey, with those inside our selection criteria
  in green.}
\label{fig.mapae}
\end{center}
\end{figure}

\begin{figure}[ht]
\begin{center}
\includegraphics[scale=0.55]{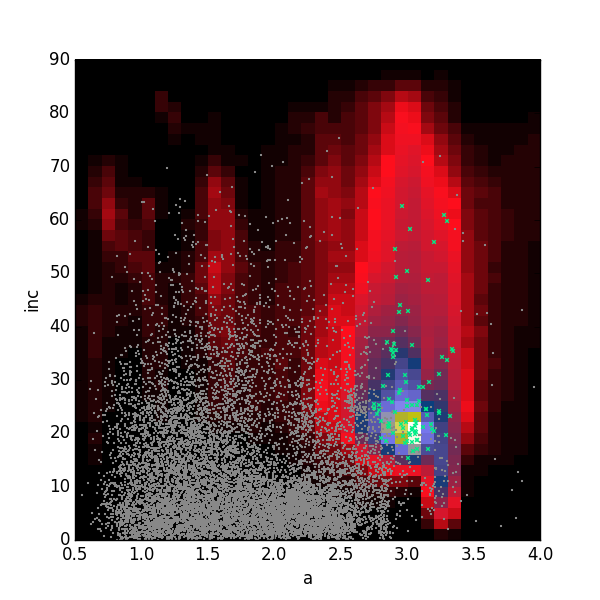}
\caption{The same as Figure~\ref{fig.mapae} but showing orbital inclination (inc) vs semimajor axis (a).}
\label{fig.mapai}
\end{center}
\end{figure}

\begin{figure}[ht]
\begin{center}
\includegraphics[scale=0.55]{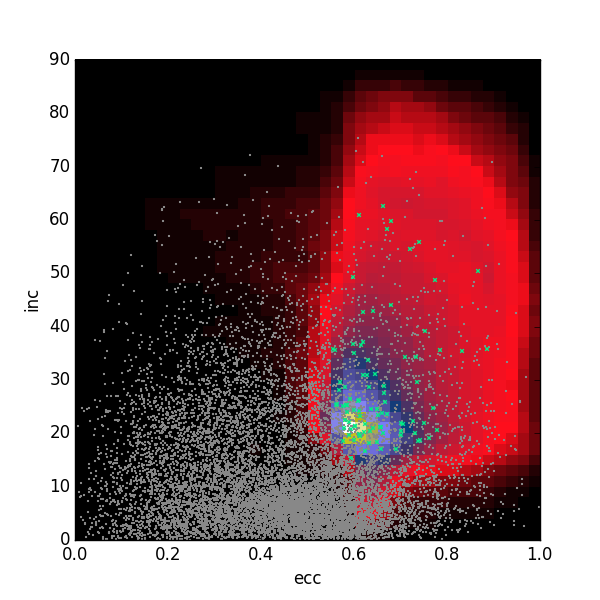}
\caption{The same as Figure~\ref{fig.mapae} but showing orbital inclination (inc) vs eccentricity (ecc).}
\label{fig.mapei}
\end{center}
\end{figure}

\begin{figure}[ht]
\begin{center}
\includegraphics[scale=0.55]{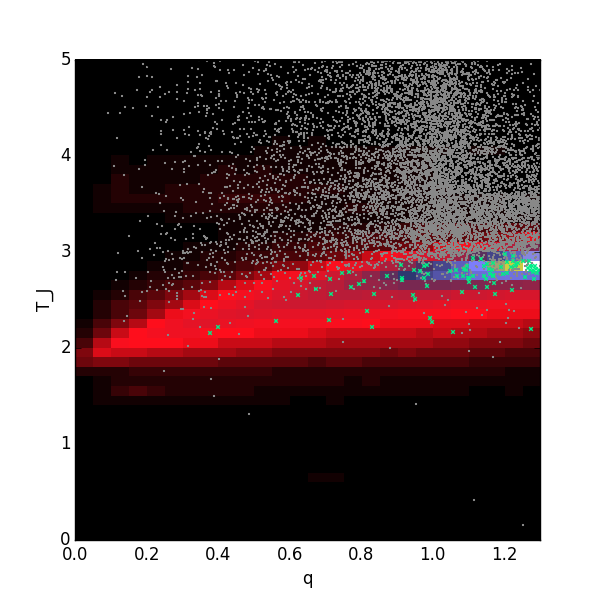}
\caption{The same as Figure~\ref{fig.mapae} but showing Tisserand
  parameter ($T_J$) vs perihelion distance (q).}
\label{fig.mapqt}
\end{center}
\end{figure}

Our simulations of family member evolution show that the primary path
for entry into the NEO population is a slow pumping of orbital
eccentricity by resonances, while semimajor axis and inclination
remain similar to the initial family values.  Simulations using the
lower initial breakup velocity show nearly identical evolution in all
respects, so we only discuss notable differences between the two
initial condition cases.  Figures~\ref{fig.mapae}-\ref{fig.mapqt} show
the probability densities of orbital parameters for Euphrosyne family
members that become NEOs, with probabilities shown as the coloring of
the bin.  At each time step each object present in the NEO population
was recorded in the appropriate bin based on its orbit to form a
probability density.  Bin values, as displayed in the figures, were
normalized such that the total probability sums to 1.  The highest
density of particles is present at a semimajor axis range slightly
smaller than the location of the family ($a\sim2.9-3.1~$AU vs
$3.08<a<3.22~$AU for the family) and at slightly lower inclinations
($inc\sim18-24^\circ$ vs $24^\circ<inc<27^\circ$ for the family), but
at significantly larger eccentricities.

As the core of the Euphrosyne family intersects the $\nu_6$ resonance,
a large fraction of our test particles ($\sim80\%$) are evacuated from
the family region and reach near-Earth space.  Thus, the Euphrosyne
family observed today is likely significantly less numerous in the
tested size ranges than the initial family formed at impact.
Unusually, smaller objects are less likely to be removed from the
family, as the larger initial velocity imparted during formation will
move them further from the region of the $\nu_6$.

Once in the NEO population, the Kozai mechanism can result in
exchanges between orbital eccentricity and orbital inclination
\citep{kozai62}, creating the fan-shaped pattern in
Figure~\ref{fig.mapei}, though the majority of particles remain near
$a\sim3~$AU, $e\sim0.6$, and $i\sim20^\circ$.  Nearly all particles in
our simulation entering the NEO population have a Tisserand parameter
$T_J<3$.  This value for the Tisserand parameter can potentially
indicate a history with the potential for gravitational interaction
with Jupiter and is often used to grade cometary orbits
\citep{levison96}.

\clearpage

These results, however, show the probability density over the entire
timescale of the simulation.  In order to compare our results to the
present-day NEO population, we need to consider the evolution of
orbital element properties of the test particles in the NEO region
over time.  Particles entering this region of phase space near the end
of the simulation will be the best analogs for real objects that may be
currently present in the NEO region.

\begin{figure}[ht]
\begin{center}
\includegraphics[scale=0.2]{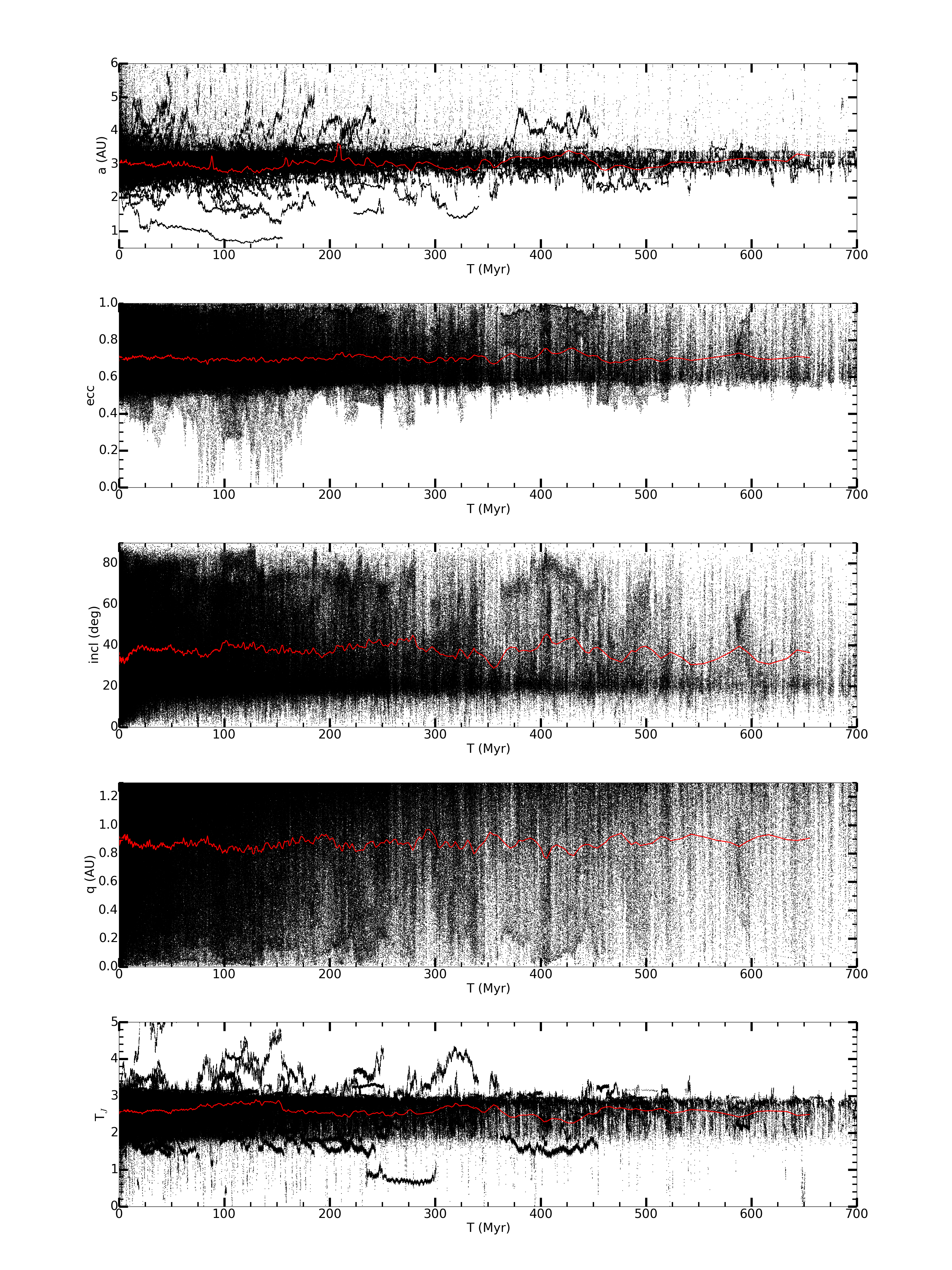}
\caption{Orbital elements for all family members in the NEO
  population, recorded every 10 kyr (points), as a function of time
  from the start of the simulation. The red line shows a running box
  mean of each element.}
\label{fig.evolve}
\end{center}
\end{figure}

In Figure~\ref{fig.evolve} we show the orbital element evolution of
all family members becoming NEOs in our simulations.  Each point
represents the instantaneous orbital elements of an object every 10
kyr.  We also show a running mean of all objects for each element.
The orbital elements of these NEOs do not change dramatically over the
timescale of our simulations, but instead show a population that is in
a rough steady-state, though with falling number density.  Most of the
particles leave the NEO population when their eccentricity approaches
$ecc=1.0$ and they impact the Sun or are ejected from the Solar
system.  We note that a handful of our test objects evolved into
Earth-like orbits, which were significantly more stable than the
majority of our population.  As such, these particles exert an
overly-large influence on the running means, and explain the
occasional large deviations of the mean from the nominal behavior of
the rest of the population.  Other than these outlier objects, the
majority of the population is found in narrow orbital ranges,
particularly for semimajor axis and Tisserand parameter, while
eccentricity and inclination both show fairly strong lower boundaries.
As the mean parameters do not change over the course of the
simulation, we can confidently use the overall probability density
maps to search for objects in the present day NEO population.

\clearpage

\begin{figure}[ht]
\begin{center}
\includegraphics[scale=0.6]{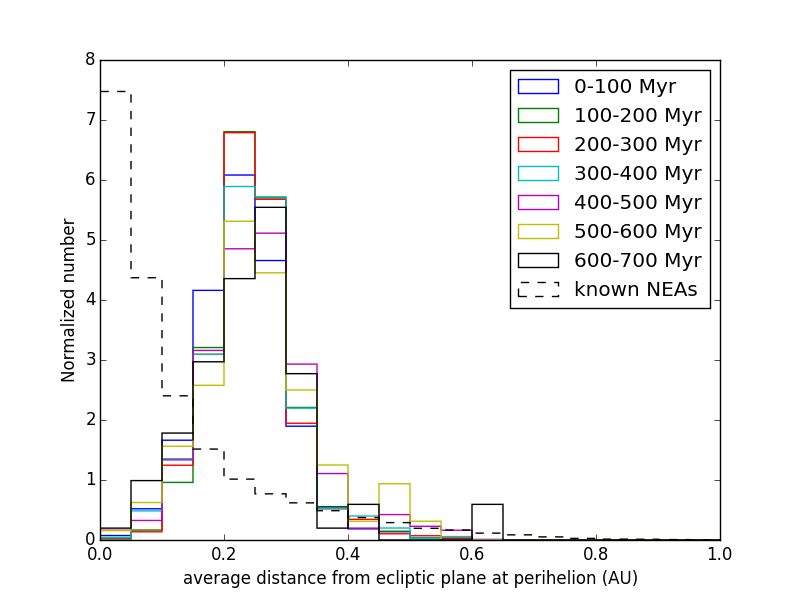}
\caption{Distribution for simulated particles of the average distance
  from the ecliptic plane at the time of perihelion. The solid lines
  show the distribution for each $100~$Myr time span of the
  simulation, while the dashed line shows the ecliptic plane distance
  for all currently known NEOs.}
\label{fig.periz}
\end{center}
\end{figure}

Euphrosyne family members that are injected into the NEO population
typically have both high eccentricities and high inclinations.  As
such, these objects will spend a majority of their orbit at a
significant distance from the plane of the ecliptic.  We investigated
the characteristic vertical distances from the plane of the ecliptic
of our test particles at perihelion ($|z_{peri}|$) as a possible
means of distinguishing objects related to the Euphrosyne family from
background NEOs.  Figure \ref{fig.periz} shows the average
$|z_{peri}|$ for all simulated objects over seven different time bins,
compared to the distribution for all known NEOs.  The $z_{peri}$
distribution for test particles does not change significantly over the
timespan of the simulation, and shows a significantly larger value
than what is observed for the known NEO population.  We can thus use
this parameter as an additional discriminant for identification of
candidate family members.

We also investigate the lifetime of family members that migrate to the
NEO population.  A longer-lived population would be expected to make
up a larger fraction of the phase space of interest than a population
that rapidly transitions out of this region. \citet{gladman00} found
that the median lifetime of NEOs was $\sim10~$Myr.  However, the
population investigated by those authors was based on all known
objects at the time, which the authors point out was biased against
the high inclination objects that dominate our sample.  Thus, we would
expect our objects to have potentially major differences from the
population tested there.

\begin{figure}[ht]
\begin{center}
\includegraphics[scale=0.6]{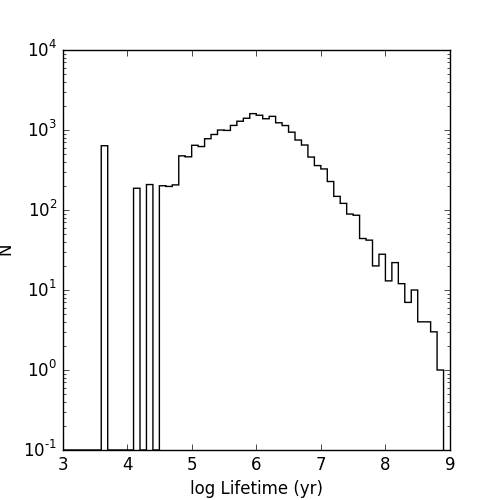}
\caption{Distribution of lifetimes within the NEO population for
  evolving family members, totaling 24000 particles.}
\label{fig.life}
\end{center}
\end{figure}

We tabulate the lifetime of our simulated objects that enter the NEO
population from their first entrance until they are removed (usually
via ejection or Solar collision as eccentricity is increased to
$e=1$). A subset of our population transitioned multiple times between
near-Earth and Mars-crossing orbits before settling into the NEO
population; however, this does not significantly impact our overall
lifetime determination.  The median lifetime and one-sigma range
determined from our simulations was $875^{+770}_{-440}~$kyr, with only
$2\%$ of test particles surviving for more than $10~$Myr.  Figure
\ref{fig.life} shows the distribution of particle lifetimes in our
simulations.  The most common end-of-life for these objects is
ejection from the Solar system or impact with the Sun when their
eccentricity is increased near $ecc=1.0$.  No difference was found in
NEO lifetime between the two different initial velocity cases tested.

\begin{figure}[ht]
\begin{center}
\includegraphics[scale=0.8]{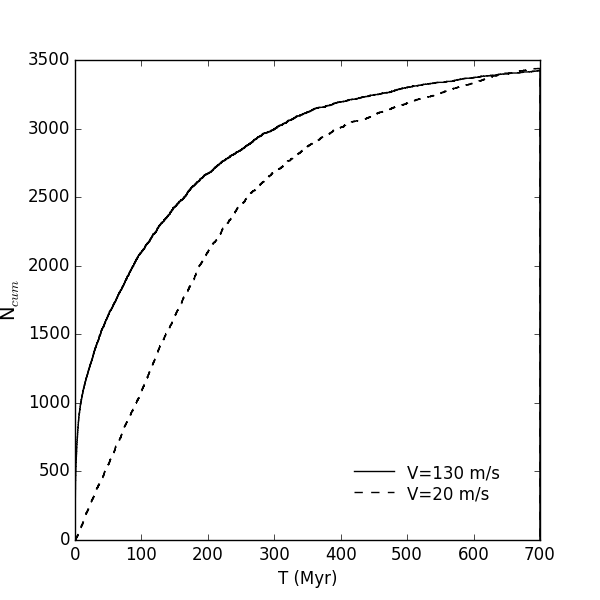}
\caption{Cumulative number of test particles from the Euphrosyne
  family injected into the NEO population over the entire timescale of
  the simulation and estimated family age.  The solid line shows
  simulations using a characteristic breakup velocity of $V=130~$m/s,
  while the dashed line shows the slower initial condition of
  $V=20~$m/s that was also tested.}
\label{fig.injTime}
\end{center}
\end{figure}

There was, however, one case of a significant difference between the two
initial breakup velocities we simulated.  In Figure~\ref{fig.injTime}
we show the cumulative number of objects transported from the family
breakup into the NEO population over the course of the simulation.
The faster initial breakup velocity resulted in a larger number of
objects being injected onto near-Earth orbits on short timescales.
However, the total population of objects transitioning into NEOs over
the time since the breakup of the family ($\sim700~$Myr) is not
significantly different. 

\clearpage

\section{Discussion}

Our simulations have shown that the breakup of the Euphrosyne family
resulted in a population of objects in near-Earth space with
characteristic orbital elements that are different from the majority
of NEOs.  The span of initial breakup velocities behaves nearly
identically in orbital element evolution, and results in comparable
contribution to the present NEO population, though the historical rate
is significantly different.  We can use the characteristic orbital
elements to search the known list of near-Earth objects for candidate
family members.

We perform multi-dimensional cuts in orbital element space based on
the highest-likelihood regions of the simulations to search for
candidate objects.  Our orbital element cuts to determine candidates
were:
\begin{itemize}
\item $2.75<a<3.35~$AU
\item $ecc>0.55$
\item $inc>15^\circ$ 
\item $inc> 90 - \frac{90~\textrm{a}}{3.4}$ 
\item $inc> 90 - \frac{90~\textrm{ecc}}{0.7}$ 
\item $1.8<T_J < 2.5 + \frac{0.5~q}{1.2}$ 
\item $|z_{peri}|>0.15~$AU
\end{itemize}

\begin{figure}[ht]
\begin{center}
\includegraphics[scale=0.6]{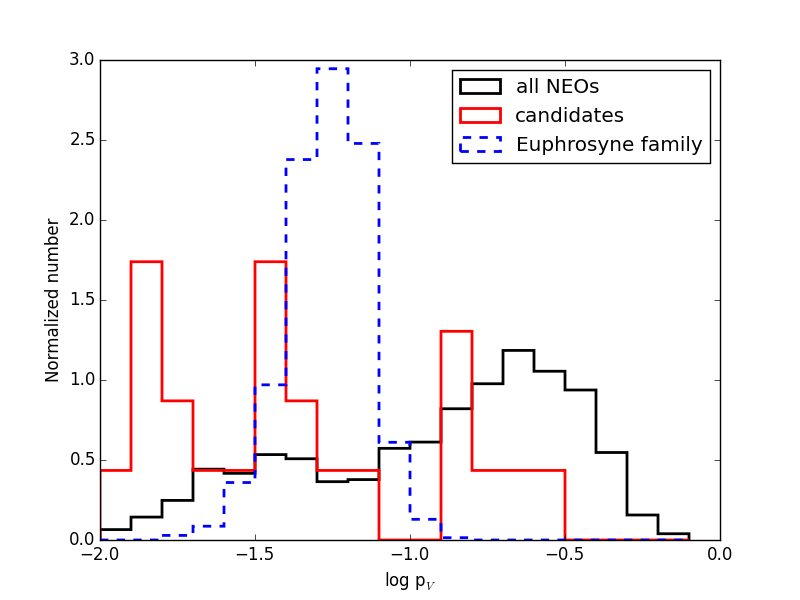}
\caption{Distribution of the geometric optical albedos ($p_V$) of all
  NEOs observed by WISE/NEOWISE (black solid, N=662), the subset of those
  NEOs identified as candidate escapees from the Euphrosyne family
  (red solid, N=23), and all Main Belt asteroids associated with the
  Euphrosyne family (blue dashed, N=1392).}
\label{fig.albs}
\end{center}
\end{figure}

After identifying the region of orbital element-space that is most
likely to be populated by escaped members of the Euphrosyne family, we
apply these limits to search all currently known NEOs.  Of the
$\sim12,000$ known NEOs to date, we find $113$ asteroids that fulfill
our selection criteria.  Of these, $23$ objects had diameters and
albedos measured by the WISE or NEOWISE missions
\citep{mainzer11nw,mainzer12pc,mainzer14restart,nugent15}, which we
list in Table~\ref{tab.candidates}.  As the WISE and NEOWISE surveys
were conducted at thermal infrared wavelengths, they are effectively
unbiased with respect to albedo, and should fairly sample both high
and low albedo objects on a given orbit.  In Figure~\ref{fig.albs} we
compare the overall distribution of albedos for all NEOs to the
distribution of albedos of our candidate objects, as well as that of
all Main Belt asteroids associated with the Euphrosyne family
\citep{masiero13}.  Low albedos are over-represented among candidate
family members when compared to the overall NEO population, however
the albedos of these objects do not uniformly match that of the
family.  Given that the residence time of Euphrosyne family members in
the NEO population is shorter than that of most NEOs, it is not
surprising that the family does not dominate even this high-likelihood
region.  We note that albedos from the post-cryo and restarted NEOWISE
missions have significant systematic uncertainties that may
artificially broaden the distribution with respect to the albedos
measured in the cryogenic portion of the NEOWISE survey.

\begin{table}[ht]
\begin{center}
\caption{Candidate Euphrosyne family members on NEO orbits having
  NEOWISE-measured physical properties.  $^\dagger$Objects discovered
  by NEOWISE.}
\vspace{1ex}
{\tiny
\noindent
\begin{tabular}{ccccccc}
\tableline
Name & a & ecc & inc & $H_V$ & $p_V$ & Diameter\\ 
MPC-packed format & AU & & deg & mag & & km \\ 
  \tableline

  O8083 & 3.324 & 0.614 & 23.341 & 16.1 & 0.126$\pm$0.072 & 2.358$\pm$0.508 \\
  O8590 & 2.911 & 0.698 & 52.363 & 16.6 & 0.020$\pm$0.008 & 4.366$\pm$0.979 \\
  f0778 & 2.910 & 0.720 & 54.500 & 18.1 & 0.048$\pm$0.032 & 1.457$\pm$0.570 \\
  f3421 & 2.889 & 0.595 & 17.408 & 18.3 & 0.023$\pm$0.018 & 1.900$\pm$0.776 \\
  f6567 & 3.120 & 0.658 & 21.107 & 17.1 & 0.036$\pm$0.011 & 2.538$\pm$0.235 \\
  f6694 & 3.059 & 0.701 & 20.525 & 17.6 & 0.031$\pm$0.007 & 2.292$\pm$0.281 \\
  f8929 & 2.893 & 0.640 & 15.204 & 17.0 & 0.137$\pm$0.027 & 1.432$\pm$0.021 \\
K00H74D & 2.919 & 0.598 & 49.262 & 18.0 & 0.163$\pm$0.029 & 0.827$\pm$0.015 \\
K03U12L & 3.291 & 0.701 & 19.737 & 17.2 & 0.200$\pm$0.051 & 1.078$\pm$0.016 \\
K09W06O & 3.086 & 0.581 & 28.760 & 17.4 & 0.034$\pm$0.008 & 2.490$\pm$0.014 \\
K09WA4F & 3.073 & 0.659 & 17.007 & 17.3 & 0.047$\pm$0.009 & 2.226$\pm$0.032 \\
K10A79G$^\dagger$ & 2.905 & 0.579 & 32.959 & 19.9 & 0.018$\pm$0.003 & 0.892$\pm$0.009 \\
K10D21M$^\dagger$ & 2.863 & 0.657 & 21.151 & 20.2 & 0.133$\pm$0.024 & 0.303$\pm$0.012 \\
K10D77H$^\dagger$ & 3.265 & 0.709 & 34.378 & 21.8 & 0.010$\pm$0.002 & 0.574$\pm$0.017 \\
K10G62X$^\dagger$ & 2.953 & 0.705 & 21.658 & 20.1 & 0.014$\pm$0.002 & 0.802$\pm$0.010 \\
K10P66R & 2.927 & 0.686 & 17.571 & 19.3 & 0.068$\pm$0.014 & 0.678$\pm$0.021 \\
K10U08B$^\dagger$ & 2.984 & 0.627 & 30.968 & 19.7 & 0.032$\pm$0.026 & 0.850$\pm$0.334 \\
K11B59N$^\dagger$ & 3.063 & 0.622 & 20.323 & 20.4 & 0.013$\pm$0.009 & 0.994$\pm$0.400 \\
K14A33A & 2.890 & 0.662 & 20.074 & 19.3 & 0.054$\pm$0.013 & 0.787$\pm$0.036 \\
K14M18Q$^\dagger$ & 2.896 & 0.600 & 35.087 & 15.6 & 0.037$\pm$0.066 & 5.272$\pm$3.496 \\
K14O01Z$^\dagger$ & 2.853 & 0.602 & 21.338 & 21.0 & 0.013$\pm$0.025 & 0.732$\pm$0.292 \\
K14R12L & 2.771 & 0.671 & 23.605 & 17.9 & 0.256$\pm$0.039 & 0.693$\pm$0.026 \\
K14X07X$^\dagger$ & 2.901 & 0.597 & 36.714 & 19.7 & 0.015$\pm$0.017 & 1.198$\pm$0.379 \\

\hline
\end{tabular}
}
\label{tab.candidates}
\end{center}
\end{table}

\clearpage

Interestingly, the low albedo objects among the candidates are in
general darker than the albedo distribution of members of the
Euphrosyne family, which correspond to the low albedo peak seen for
the outer Main Belt.  Caution is needed in interpreting this result,
as our sample does suffer from small-number statistics and detection
biases, as well as potential contamination from other source regions.
However, if these candidates are indeed escapees from the Euphrosyne
family we would need to identify a plausible explanation for the
difference in albedo from the family.  A potential avenue is the
investigation of space weathering effects on low albedo asteroids or
changes in optical properties due to changing surface regolith size.
Recent work comparing colors and albedos of C-type families has
indicated that low albedo asteroids become redder and darker with
time, but that smaller C-type asteroids may have lower albedos than
larger ones \citep{kaluna15}.  Investigation of these effects would
benefit from further characterization of the physical properties of
NEOs, particularly albedos and spectra.

We also used the limits derived from our simulations to search the NEO
population model of \citet{greenstreet12}.  This model tabulates the
fractional representation of a debiased NEO population across
near-Earth space, tracing bins back to one of five NEO source regions.
Using the above limits (except for the $z_{peri}$ limit as this
information is not available in the Greenstreet model), we find that
less than $4\%$ of simulated NEOs are predicted to be found in our
region of interest, and of those only $3\%$ originated from the
$\nu_6$ resonance.  We note that the simulations of
\citet{greenstreet12} follow \citet{bottke02} and only consider the
region of the Main Belt near the $\nu_6$ resonance at semimajor axes
$a<2.5~$AU.  The majority ($54\%$) of objects in this region are
predicted to evolve into the NEO region from the Jupiter-family comets
(JFCs), while $25\%$ originated in the outer Main Belt.  Again
following \citet{bottke02}, the outer Main Belt region tested by
\citet{greenstreet12} was limited to inclinations $i<15^\circ$,
significantly below our the initial conditions for our simulations.
\citet{bottke02} do include this dynamical region in their MB2
population, however they state that it is only a minor contributor to
the NEO population.  We note that the statistically determined
percentage of objects in this region originating from JFCs matches the
estimates of the cometary contribution to the NEOs with $T_J<3$ by
\citet{demeo08} of $\sim54\%\pm10\%$ based on observed physical
properties.

Given the simulations of \citet{greenstreet12}, it is possible that
our candidate list is significantly contaminated by inactive JFCs.
\citet{lamy04} show that the albedos of bare JFC nuclei typically
range from $0.02-0.04$, which is lower than the albedos of Euphrosyne
family members ($p_V=0.06\pm0.02$).  Using a thermal infrared survey
conducted with the Spitzer Space Telescope, \citet{fernandez13} found
that the median diameter of JFCs is $\sim2-5~$km, similar in size to
the majority of the test particles in our simulations, though with a
cumulative size frequency distribution (SFD) slope of $\alpha =
-1.9\pm0.2$ where the SFD is described as $N_{>D} \propto D^{\alpha}$.
Preliminary NEOWISE results on a larger sample of comet nuclei confirm
this shallow SFD \citep{bauer15}.  This value is significantly
shallower than the present day Euphrosyne family, which has an SFD
slope of $\alpha=-4.40\pm0.05$ \citep{masiero13}, our initialized
family SFD from \citet{carruba14} of $\alpha=-3.6$, or the SFD of the
near-Earth object population created in our simulations
($\alpha=-3.81\pm0.02$).  Thus JFCs may explain part of lowest albedo
objects in our candidate albedo distribution, however given the
shallower size distribution we would expect the relative contribution
of JFCs to decrease for smaller objects.

We also compare our candidate source region to the objects expected
from the debiased NEO population of \citet{mainzer12subpop}.  Given
the semimajor axis limits listed above, the region of interest falls
entirely within the Amor subpopulation (objects with perihelia
$1.0<q<1.3~$AU).  Applying the cuts derived from our simulations to
the debiased population shows that we should expect approximately
$2.3\pm 0.2\%$ of all Amor NEOs and $0.8\pm 0.2\%$ of all NEOs to have
orbital elements consistent with family member candidates.  Low albedo
NEOs ($p_V<0.1$) make up $39.1\pm 0.7\%$ of all NEOs in the debiased
population and $41.7\pm 0.7\%$ of the debiased Amor population.  If
our candidate objects all had low albedos they could account for the
majority of the excess of dark asteroids in the Amor population when
compared to the general NEO population.  As seen in
Fig~\ref{fig.albs}, however, the candidate objects with measured
albedos show some contamination from high albedo asteroids, thus we
consider it unlikely that refugees from the Euphrosyne family explain
the difference between the Amor and NEO albedo distributions completely.

\citet{laspina04} showed that NEOs have spin states that are
preferentially retrograde, a hallmark of their evolution under
Yarkovsky driving them inward toward the $\nu_6$ resonance where it
intersects the inner Main Belt.  As the Euphrosyne family has
overlapped the $\nu_6$ resonance from formation, we would not expect a
similar selection effect in the rotation poles by Yarkovsky
mobilization, and instead would expect family members to show equal
populations of prograde and retrograde rotators.  None of the objects
in our candidate list have published rotation poles in the Asteroid
Lightcurve Database \citep{warner09}, however future surveys may
provide insight into the rotation properties of this population.

\section{Conclusions}

The Euphrosyne family is one of the largest low albedo families known,
and its unique location in the outer Main Belt overlapping the $\nu_6$
resonance means that its contribution to the NEO population has a
unique signature.  Our simulations of asteroid orbital evolution over
the age of the family allow us to determine the best region of orbital
element-space to search for candidate family members among the NEOs.
The most probable region is found near the semimajor axis of the
family, at large inclinations and very large eccentricities.  This
results in a Jupiter Tisserand parameter similar to what is observed
for Jupiter-family comets.  We also find that candidate family members
typically have large vertical distances from the ecliptic at
perihelion, as opposed to the majority of known NEOs that tend to
reach perihelion close to the ecliptic plane (although we note this is
almost undoubtedly biased by observational selection effects favoring
detection of NEOs close to the Earth and thus near the ecliptic
plane).  

Using the limits derived here, we identify the currently known NEOs
that are most likely to have evolved from the Euphrosyne breakup into
near-Earth orbits, $\sim1\%$ of total population.  Approximately
$20\%$ of candidates have diameter and albedo measurements in at least
one phase of the NEOWISE mission, allowing direct comparison to the
Euphrosyne family, and the overall NEO population.  We find that
candidates are preferentially lower in albedo than the overall NEO
population.  Many candidates also are darker than the Euphrosyne
family, which may be indicative of the contribution of dead comets
from the JFC population into the NEOs in this region of orbital
element space.  Followup observations of colors and spectral
properties would help differentiate dormant/dead comets from
Euphrosyne family members.  As the Euphrosyne family has a steeper
size distribution than JFCs, future surveys of NEOs to smaller sizes
(especially thermal infrared surveys that are not biased against low
albedo objects) would have a high likelihood of discovering a
significant number of escapees from this family.

\section*{Acknowledgments}

This research was carried out at the Jet Propulsion Laboratory,
California Institute of Technology, under a contract with the National
Aeronautics and Space Administration.  JM was funded by a NASA
Planetary Geology and Geophysics grant, and through the JPL internal
Research and Technology Development program.  VC was supported by the
FAPESP grant 2014/06762-2.  The authors would like to thank UNESP,
CAPES, AAB, and FAPESP for supporting the 2014 Small Bodies Dynamics
conference in Ubatuba, Brazil, which inspired this work.  The JPL
High-Performance Computing Facility used for our simulations is
supported by the JPL Office of the CIO. This publication makes use of
data products from the Wide-field Infrared Survey Explorer, which is a
joint project of the University of California, Los Angeles, and the
Jet Propulsion Laboratory/California Institute of Technology, funded
by the National Aeronautics and Space Administration.  This
publication also makes use of data products from NEOWISE, which is a
project of the Jet Propulsion Laboratory/California Institute of
Technology, funded by the Planetary Science Division of the National
Aeronautics and Space Administration.

\end{document}